\documentclass[english,preprint,aps]{revtex4-1}
\usepackage[LGR,T1]{fontenc}
\usepackage[latin9]{inputenc}
\usepackage{color}
\usepackage{textcomp}
\usepackage{amsmath}
\usepackage{amssymb}
\usepackage{graphicx}

\makeatletter

\newcommand{\lyxmathsym}[1]{\ifmmode\begingroup\def\b@ld{bold}
  \text{\ifx\math@version\b@ld\bfseries\fi#1}\endgroup\else#1\fi}

\makeatother

\usepackage{babel}
\begin{document}
\title{Seismic 1/f Fluctuations from Amplitude Modulated Earth's Free Oscillation}
\author{Akika Nakamichi}
\email{nakamichi@cc.kyoto-su.ac.jp}

\affiliation{General Education, Kyoto Sangyo University ~\\
Motoyama Kamigamo Kita-ku, Kyoto 603-8555 Japan }
\author{Manaya Matsui}
\email{m-matsui1@neitec-grp.com}

\affiliation{Meitec Corporation ~\\
1-1-10 Ueno, Taito-ku Tokyo 110-0005 Japan }
\author{Masahiro Morikawa}
\email{hiro@phys.ocha.ac.jp}

\affiliation{Department of Physics, Ochanomizu University ~\\
2-1-1 Otsuka, Bunkyo, Tokyo 112-8610, Japan }
\begin{abstract}
We first report that the seismic time-sequence data from around the
world, excluding major earthquakes, consistently yield the power spectral
density inversely proportional to the frequency f. This is the 1/f
fluctuation that appears ubiquitously in nature. We investigate the
origin of this 1/f fluctuation based on our recent proposal: 1/f noise
is amplitude modulation and demodulation. We hypothesize that the
amplitude modulation is linked to resonance with Earth's Free Oscillations
(EFO), with demodulation occurring during fault ruptures. We provide
partial validation of this hypothesis through an analysis of EFO eigenmodes.
Additionally, we outline potential methods for the future verification
of our theory relating 1/f fluctuations to EFO. 
\end{abstract}
\maketitle

\section{Introduction}

Seismic activities are complex phenomena caused by multiple factors
with highly nonlinear interactions. Despite this complexity, seismic
activities exhibit certain universal scaling laws, such as the Gutenberg-Richter
(GR) law\citep{Gutenberg1944} and the Omori law\citep{Omori1894}.
Both laws can be represented as power laws and describe the essence
of earthquake occurrence. Recently, yet another universal law has
been reported to describe the inter-occurrence time distribution of
earthquakes by the Weibull distribution function\citep{Tanaka2017,Hatano2017}.
These laws focus on the local property: statistics of the individual
seismic events or adjacent events or clustered events. This paper
represents a departure from local descriptions, aiming instead to
globally characterize seismic events through an analysis of the entire
time series of worldwide seismic activities. 

We try power spectral density (PSD) analysis of a long period, including
all recorded earthquakes in the whole Earth, using the USGS dataset
\citep{USGS2023}. However, this attempt fails; all appear to be random.
Next, we try the same, excluding giant earthquakes. Then a clear power
law, with an index of about $-1$, appears in the low-frequency range.
This is the 1/f fluctuation (PSD with a power index from $-1.5$ to
$-0.5$) often observed in various fields of nature. The power law
appears most clearly if we entirely disregard the energy information
and analyze the time series of the seismic activity occurrence. This
finding suggests that seismic 1/f fluctuations are predominantly related
to low-energy phenomena, likely triggered by minor energy sources. 

Incidentally, we are not the first to find 1/f fluctuations in seismic
activities; authors of \citep{Bittner1996,Lapenna2000,Telescaa2001}
analyzed earthquakes in the Italian district and found 1/f fluctuations
in their time sequence.

We want to reveal the origin of the seismic 1/f fluctuation in this
paper. We've recently proposed a simple model of 1/f fluctuations
from the beat of many waves with accumulating frequencies \citep{Morikawa2023}.
This is an amplitude modulation. In particular, resonance can naturally
yield the accumulation of wave frequencies \citep{Morikawa2023}.
What on Earth is resonating?

We speculate that the whole lithosphere is resonating, \textit{i.e.,}
Earth Free Oscillation (EFO) \citep{Alterman1959,Benioff1961,Ness1961,Alsop1961,Nawa1998,Woodhouse2007,Masters1995}.
Thus, we collect so far calculated eigenfrequencies of EFO and construct
the PSD, expecting the 1/f fluctuation to appear.

\section{1/f fluctuations in global earthquakes\label{sec:Power-spectrum-of}}

\begin{figure}
\includegraphics[width=12cm]{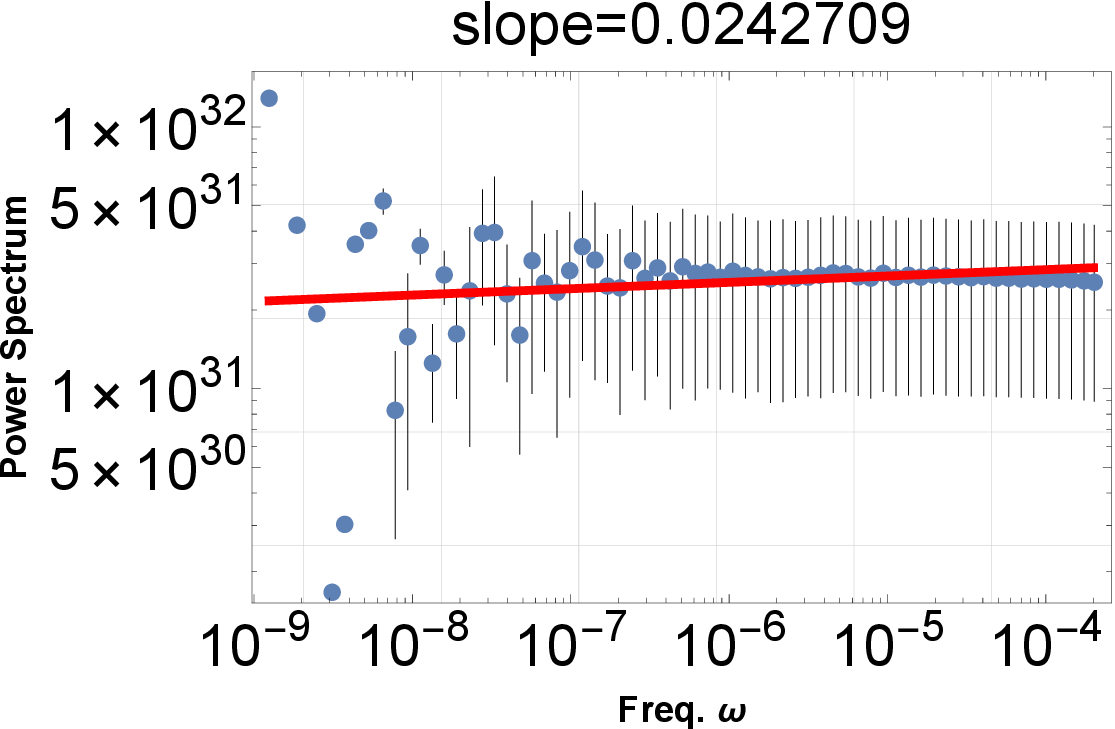}\caption{(Color online) Power spectral density (PSD) of the energy time sequence
of all the world earthquake data fifty years 1972-2022 (USGS). The
time is measured in seconds, and the frequency unit is Hz. Before
analysis, the data are homogenized over time. Specifically, the total
time interval is divided into equal segments corresponding to the
number of events, and the energy of events within each segment is
allocated accordingly. Same for all PSD analyses below. \protect
\protect \protect \\
 The present result is entirely random, as shown by the flat red line
that fits the data points. }
\label{fig1} 
\end{figure}

Thanks to the 50 years of compiling the world earthquake data at USGS,
we can effectively analyze the Fourier power spectrum of the time
sequence. First, we use all the global data of earthquakes with a
magnitude greater than $3.5$ and prepare the entire time sequence
of the energy released by each earthquake event $E(t)=10^{4.8+1.5M}\mathrm{Joule}$,
where M is the magnitude of the earthquake at time $t$. Then the
power spectral density (PSD) for this energy time sequence becomes
entirely random, as shown in Fig.\ref{fig1}.

\begin{figure}
\includegraphics[width=12cm]{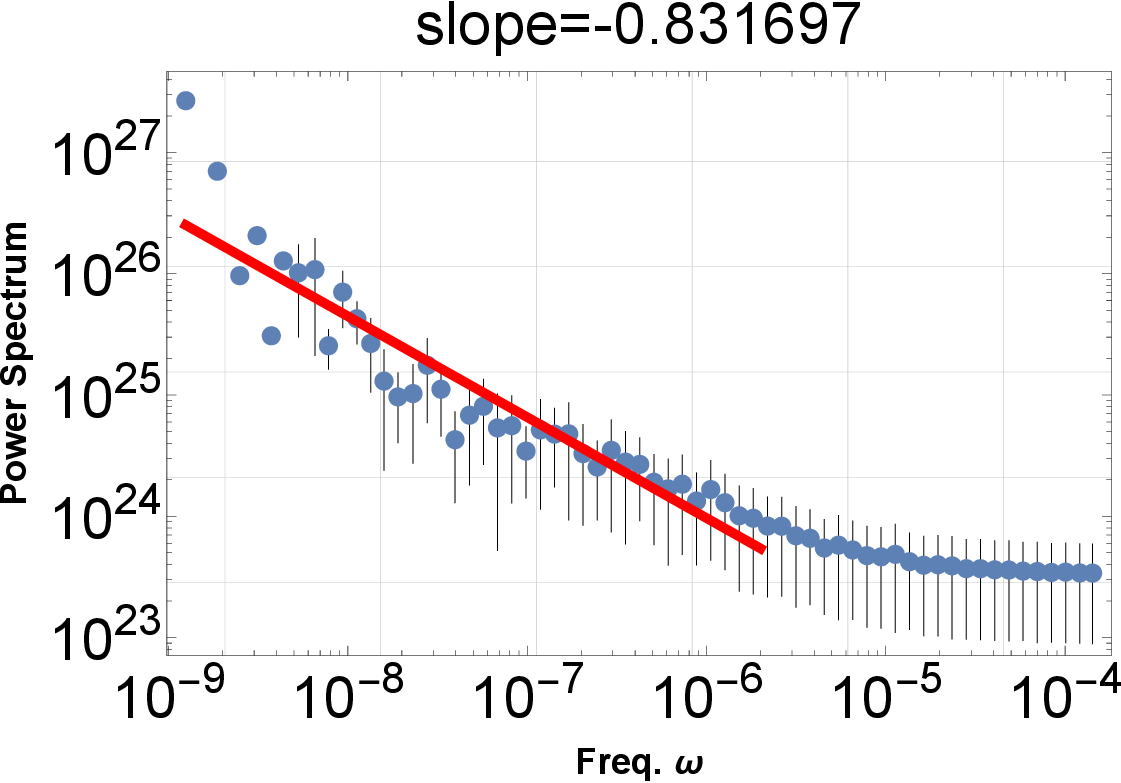}\caption{(Color online) Same as Fig.\ref{fig1}, but the magnitude range is
limited up to 5. The PSD shows an obvious 1/f fluctuation with an
index of -0.83.}
\label{fig2} 
\end{figure}

Subsequently, we repeated this analysis while restricting the magnitude
range from 3.5 to 5, thereby excluding larger earthquakes. Then the
low-frequency signal in PSD appears, as shown in Fig. \ref{fig2},
as a power-law with an index of -0.83, a typical 1/f fluctuation.
Interestingly, the whole global data appears to have long temporal
memory of 50 years. For the sake of completeness, we calculated PSD
for the earthquake magnitude greater than 5 and obtained a completely
random result similar to Fig.\ref{fig1}. These facts indicate that
seismic 1/f fluctuations are low energy phenomena.

\begin{figure}
\includegraphics[width=12cm]{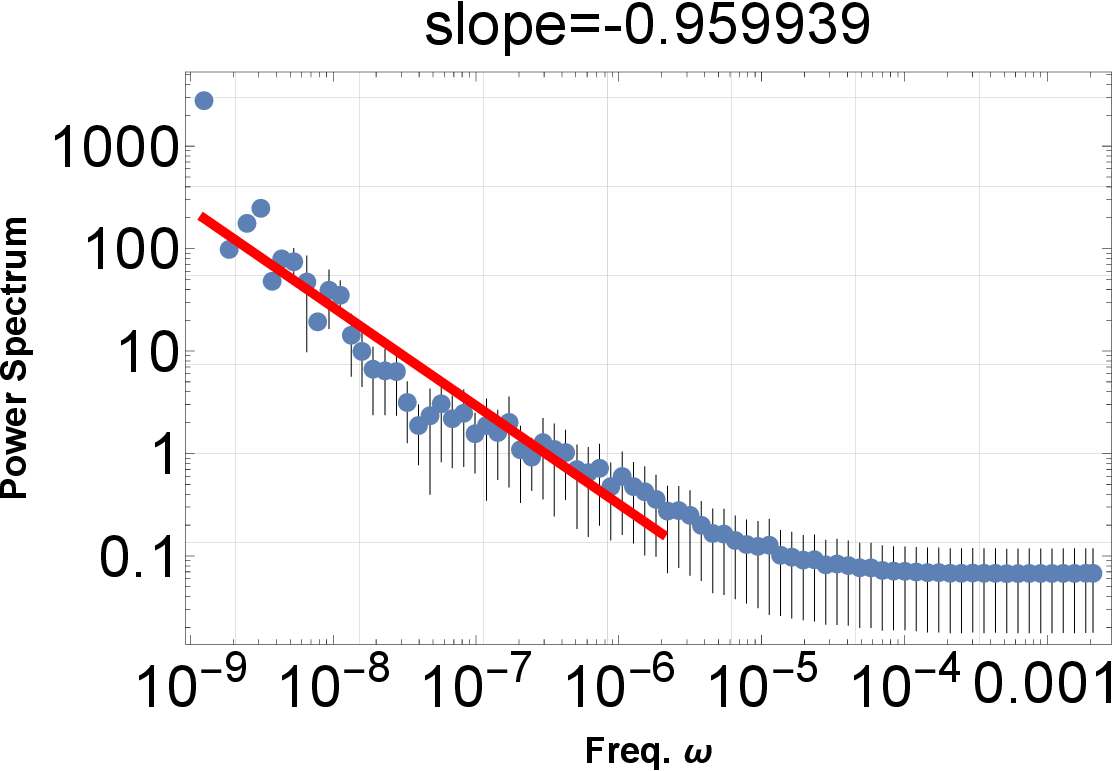} \caption{(Color online) Same as Fig.\ref{fig1}, but the energy information
is removed; we set the energy value to one for each data. The PSD
shows the power behavior with an index of -0.96, a typical 1/f fluctuation. }

\label{fig3} 
\end{figure}

Moreover, upon removing all specific energy values and standardizing
each event's energy as one in the time sequence, the PSD exhibits
a 1/f fluctuation with a power index of -0.96, as depicted in Fig.\ref{fig3}.
These facts indicate that the seismic 1/f fluctuation cannot be caused
by the self-similar fractal structure from the small to large energy
scales. On the contrary, the facts indicate that the seismic 1/f fluctuations
are low-energy phenomena, probably triggered by a tiny energy source.

Remarkably, the 1/f fluctuations with the power index $-0.8\sim\lyxmathsym{\textendash}1.0$
are observed in the wide range of frequencies corresponding to timescales
from about a week ($2\times10^{-6}$ Hz) to several decades ($10^{-9}$
Hz). This raises the question: what mechanism produces a coherent
structure over such a broad time range, spanning approximately three
orders of magnitude, with only a low-energy trigger? 

Based on the aforementioned findings, the earthquake appears to be
two consecutive processes: a) the gradual accrual of seismic energy
in the form of accumulating stress within the fault, and b) a subsequent
trigger that disrupts the locked fault, initiating a sudden discharge
of energy. It is evident from the preceding discussions that the 1/f
fluctuation is intricately linked to the second component, b).

The first process, a), involves the progressive buildup of elastic
strain energy within the fault due to the random accumulation of energy
around the locked fault surface. This phenomenon may be well described
by the principles outlined in the theory of self-organized criticality.

As for the second process, b), the final trigger introduces a small
amount of energy, causing the cumulative stress energy to surpass
the frictional threshold and inducing the fault to slip. This minute
trigger could potentially be attributed to perpetual fluctuations
in the lithosphere, and it exhibits properties of 1/f fluctuation.

In light of these findings, our subsequent analysis will concentrate
on exploring the second process, b), in greater detail. 

\section{Amplitude modulation and wave beats\label{sec:Amplitude-modulation-and}}

In our recent work\citep{Morikawa2023}, we proposed amplitude modulation
as a potential origin for the observed 1/f fluctuation. This theory
hinges on the beat of many waves with accumulating frequencies. This
perspective diverges from the prevalent theories rooted in self-organized
criticality or multifractal geometry. For the successful production
of beats or amplitude modulation for 1/f fluctuation, we need a resonance
in which many adjacent modes with accumulating frequencies exist.

We hypothesize that Earth Free Oscillation (EFO) serves as the resonant
mode essential for generating 1/f fluctuations, which is always excited
over the entire lithosphere\citep{Alterman1959,Benioff1961,Ness1961,Alsop1961}.
Specifically, surface wave modes -- both toroidal and spheroidal
-- exhibit accumulating eigenfrequencies as the angular index $l$
decreases\citep{Woodhouse2007}. We show that this accumulation successfully
yields a 1/f power spectral density.

The excited EFO waves are continuous and always propagating in the
lithosphere. An earthquake event is triggered when the energy of these
EFO waves surpasses a threshold, providing the final impetus for a
fault slip. In this process, the continuous EFO waves yield discrete
events of earthquakes. Therefore, the 1/f fluctuation property should
arise in this thresholded discrete sequence. 

The requirement of thresholding to observe 1/f fluctuations may serve
as a critical verification point for our amplitude modulation (AM)
proposal. According to our proposal, 1/f fluctuation is encoded as
the amplitude modulation in relatively high-frequency waves. The 1/f
fluctuation property only arises after some demodulation process;
the positive and negative parts of the original fluctuating wave in
relatively high frequency, including 1/f modulation, cancel each other
out \citep{Morikawa2023}. 

Subsequently, we will analyze EFO waves to demonstrate that 1/f fluctuations
manifest only post-thresholding, not within the original wave form. 

\section{Resonating Earth's Free Oscillation \label{sec:Resonator-Earth-Free}}

In this section, we investigate the hypothesis that Earth Free Oscillation
(EFO) is a key trigger for 1/f fluctuations observed in seismic activities.
In particular, we focus on how the EFO eigenmodes contribute to the
accumulation of frequencies and provide low-frequency signals through
amplitude modulation mechanisms.

The small displacement $u(t,r,\theta,\phi)$ of the elastic Earth
from the equilibrium position obeys the balance of forces, Hooke's
law, and the Poisson equations 
\begin{equation}
\rho\ddot{u}=-\nabla p-\rho\nabla\phi_{g},\:p=-\kappa\nabla u,\:\triangle\phi_{g}=4\pi G\rho
\end{equation}
that reduce to the wave equation 
\begin{equation}
\rho\ddot{u}=\kappa\triangle u-\rho\nabla\phi_{g},
\end{equation}
where $p,\rho,\kappa,G,\phi_{g}$ are the pressure, mass density,
bulk modulus, gravitational constant, and gravitational potential,
respectively. The stationary solution $u(t,r,\theta,\phi)=v(r,\theta,\phi)e^{-i\omega t}$
yields the eigenvalue equation. The variable separation method in
the spherical coordinate system yields the solution of the form 
\begin{equation}
u(t,r,\theta,\phi)=R_{n,l,m}(r)Y_{l}^{m}(\theta,\phi)e^{-i\omega_{n,l,m}t},
\end{equation}
where $Y_{l}^{m}(\theta,\phi)$ is the spherical harmonics and the
modes are labeled by $n=0,1,2,...$, $l=0,1,2...$, and $-l\leqq m\leqq l$.
The modes are classified into spherical and toroidal modes, stretching
vibration and torsional vibration, respectively. All the parameters
depend on the details of the Earth's interior, and solving the eigenvalue
equation is a complicated task.

Thanks to the compilation of observational data and numerical calculations
of EFO eigenfrequencies by many researchers so far, we are ready to
calculate the amplitude modulation. Initially, we utilize numerically
calculated eigenfrequency data derived from the Earth model PREM {[}16{]},
as detailed in Table 1. This table contains useful information about
many modes, observed frequencies, and model frequencies,... A characteristic
feature of the modes is the accumulation of frequencies towards smaller
$l$ for each $n$ in both spherical and toroidal modes. This property
is crucial for observing the 1/f fluctuations. We randomly superimpose
all the sinusoidal waves with frequencies from the lowest $309.28\mu Hz$
($_{0}S_{2}$) up to $9994.61\mu Hz$ ($_{14}S_{19}$) where $_{n}S_{l}$
means the spheroidal mode with overtone index $n$ and harmonic degree
$l$. We also include all the toroidal modes $_{n}T_{l}$ within this
frequency range. The frequency degenerates with respect to the azimuthal
order number $m$ ($-l\leq m\leq l$) in the present static spherical
model. The wave mode superposition becomes 
\begin{equation}
\Phi(t)=\sum_{k=1}^{N}\xi_{k}sin(2\ensuremath{\pi}\Omega_{k}t),\label{eq Phi}
\end{equation}
where $\xi_{k}$ is the random variable in the range $[0,1]$, and
$N=1158$ is the total number of eigenfrequencies in the above range.
Then, we Fourier analyze the power spectral density (PSD) for the
time series of the absolute value $\left|\Phi(t)\right|$.We have
no signal in the low-frequency domain if we calculate PSD for the
bare $\Phi(t)$. Since the 1/f fluctuation signal is modulated in
our model, some demodulation process is needed; taking the absolute
value is a typical demodulation. This process is needed to extract
1/f fluctuations in the PSD analysis, and the actual demodulation
will be inherent in the system. This point will be discussed later.

\begin{figure}
\includegraphics[width=12cm]{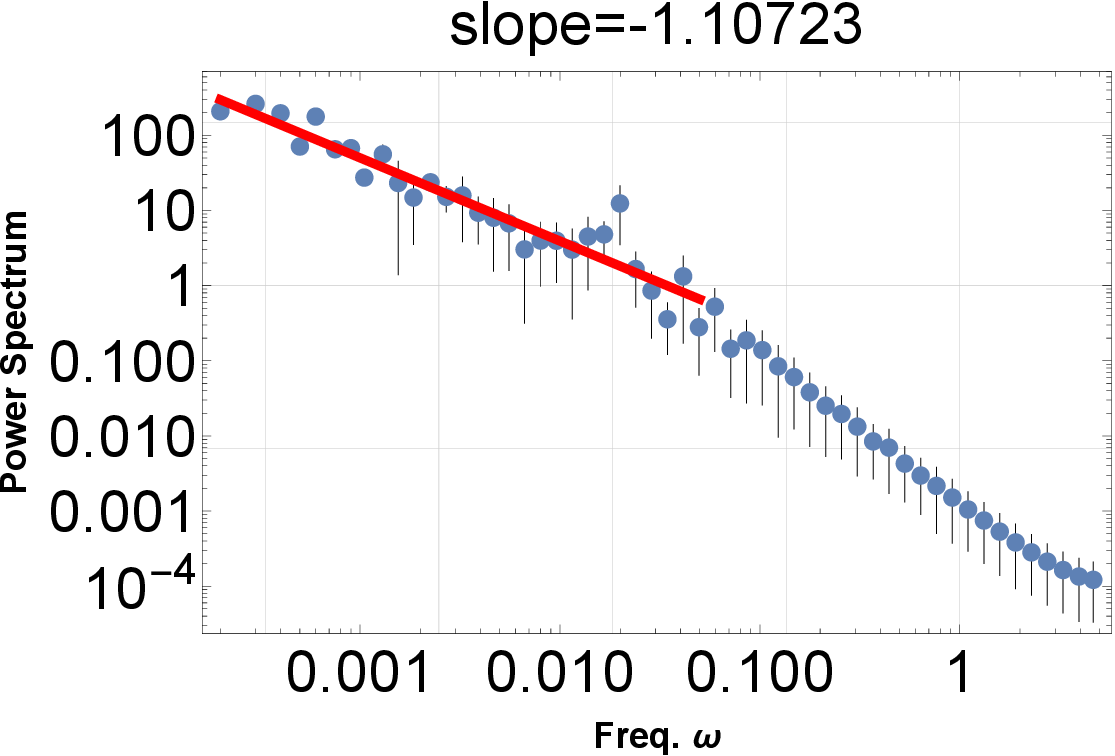}\caption{(Color online) The PSD of the absolute value of the time sequence
Eq.\ref{eq Phi} $\left|\Phi(t)\right|$. $\Phi(t)$ is the superposition
of waves with the first $N=1158$ frequencies of the EFO from the
lowest, with random amplitude. Each mode is labeled by $n,l$, and
$m$ is degenerate. This shows a typical 1/f fluctuation with an index
of $-1.1$ for three decades. }
\label{fig4} 
\end{figure}

As the result of the PSD analysis, we obtain a power-law with an index
of about -1.0 within the low-frequency range of $10^{-4}-10^{-2}Hz$
as shown in Fig.\ref{fig4}. On the other hand, the observed seismic
activity is in the range of $10^{-9}-10^{-6}Hz$; far lower than this
analysis. This gap can be filled by considering more realistic fine
structures of the eigenstates and more resonances. We now proceed
to such an analysis.

So far, our analysis of the resonance is not complete. For example,
a) each eigenmode accompanies a resonance curve, and many more accumulating
modes are associated with each mode, b) so far degenerate modes in
the azimuthal order number $m$ should yield fine structure around
each principal frequency labeled by $n,l$. This degeneracy in $m$
is removed by the Earth's non-spherical symmetry or the Earth's rotation.
In this paper, we analyze typical representative modes for both cases
a) and b) to show that the fine structure extends the 1/f fluctuation
power to a much lower frequency range. A complete analysis will be
reported in our future publications.

To refine the PSD, we account for two critical factors: a) each eigenfrequency,
denoted by $n,l$, possesses a finite width, and b) the degeneracy
in m is resolved due to the Earth's rotation.

a) The resonant modes are expressed by the Lorentzian distribution,
\begin{equation}
R[\omega]=\frac{1}{\left(\frac{\kappa}{2}\right)^{2}+\left(\omega-\omega_{0}\right){}^{2}},\label{eq:R}
\end{equation}
where $\omega_{0}$ is the fiducial resonance frequency and $\kappa$
characterizes the sharpness of the resonance. This function represents
the frequency distribution density associated with the fiducial frequency
$\omega_{0}$. Then the inverse function (tangent) of the cumulative
distribution function (hyperbolic tangent) generates this distribution
from the Poisson random field.

b) The Earth's rotation resolves the degeneracy in $m$ by breaking
the spherical symmetry of the system. The details are complicated,
but the rough estimate is given by the resolved frequency \citep{Backus1961,Gilbert1965}
in the lowest perturbation in $\Omega_{\mathrm{EarthRotation}}/\omega_{nl}(\ll1)$,
\begin{equation}
\omega_{nlm}=\omega_{nl}+\frac{m}{l(l+1)}\Omega_{\mathrm{EarthRotation}},\label{eq:Omega}
\end{equation}
where $\omega_{nl}$ is the degenerate eigenfrequency and $\Omega_{\mathrm{EarthRotation}}=1.16\times10^{-5}Hz$
is the frequency associated with the Earth's rotation. The coefficient
of $\Omega_{\mathrm{EarthRotation}}$ is exact for the torsional modes
but is approximate for the spheroidal modes.

These effects are processed as follows. We first construct wave data
superposing $N$ sinusoidal waves with eigenfrequencies after removing
the degeneracy in $m$. We further superimpose $M$ resonant waves
of frequencies close to the fiducial frequency according to the distribution
Eq.(\ref{eq:R}). The fully superposed wave becomes

\begin{equation}
\Phi(t)=\sum_{n=1}^{N}\sum_{i=1}^{M}\sin\left(2\pi(1+c\tan(\xi_{i}))\Omega_{n}t\right),\label{eq:Phi2}
\end{equation}
where the parameter $c=\kappa/\Omega_{n}$ represents the relative
line width for each eigenfrequency. The random variable $\xi_{i}$,
running in the range $[0,\pi/2]$, generates the frequency distribution
by $R(\omega)$ in Eq.(\ref{eq:R}). The parameter $c$ actually depends
on each $n$, but according to the table in \citep{Masters1995},
$c$ turns out to be of order $0.01$ and we use this constant value:
$c=0.01$. We used the catalog \citep{Chung2017}, limiting the numbers
to $M=1000$ and $N=100$.

\begin{figure}
\includegraphics[width=12cm]{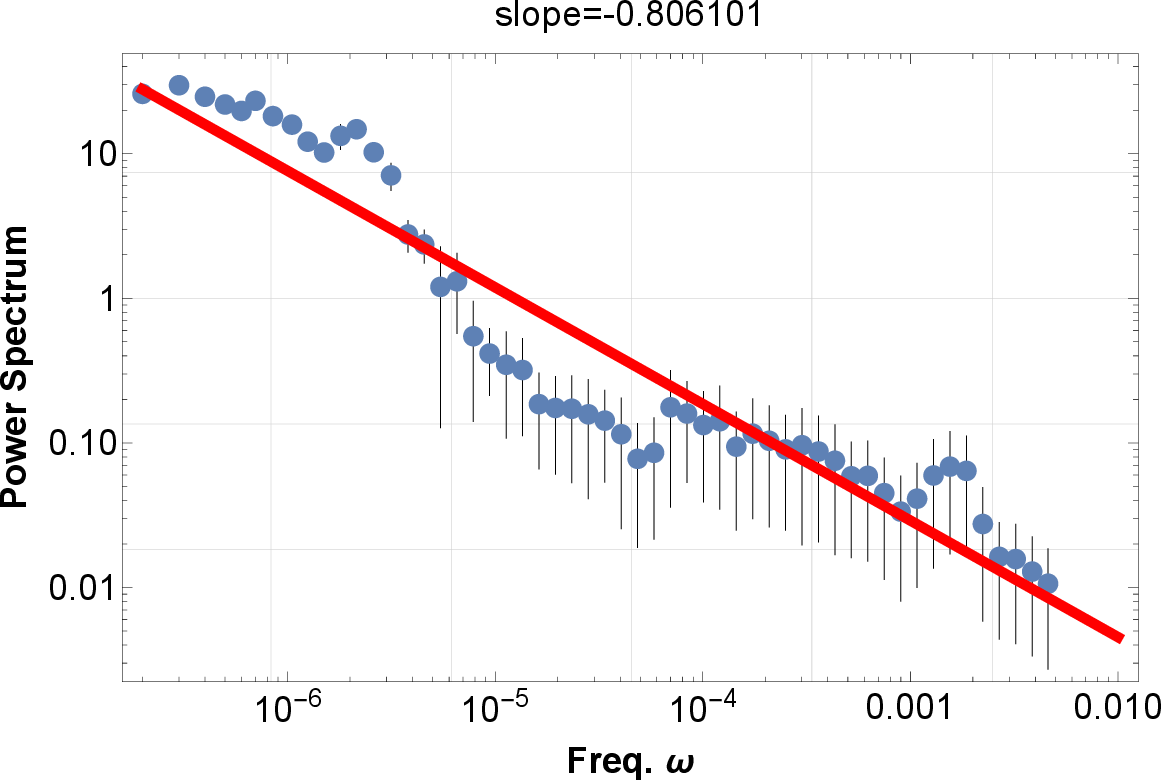}\caption{(Color online) Same as Fig.\ref{fig4}, but including resonant modes
and fine eigenmodes after resolving the degeneracy in $m$. We first
construct the data $\Phi(t)$ as a superposition of sinusoidal waves
with $100$ frequencies from the lowest and $N=1000$ Lorentzian distributed
modes. The latter is randomly generated according to Eq.\ref{eq:R}.
The graph shows the PSD of the thresholded value of the time sequence
Eq.\ref{eq:Phi2} $\left|\Phi(t)\right|$. The threshold is set to
select the data points $\left|\Phi(t)\right|$that are greater than
twice the mean. This graph shows nearly the 1/f fluctuation with an
index of $-0.81$ for over four decades. The PSD of other thresholds
and different sample sizes also yield similar 1/f fluctuations. }

\label{fig5} 
\end{figure}

As before, the PSD of the bare $\Phi(t)$ gives no signal in the low-frequency
region. However, the absolute value $\left|\Phi(t)\right|$ or arbitrarily
set threshold data gives 1/f fluctuations (details are in the caption
of Fig.\ref{fig5}). These square operation and thresholding work
as a demodulation of the original signal. In this way, the 1/f fluctuation
appaears only after demodulation and is quite robust. Figure \ref{fig5}
shows the PSD of the thresholded data. It shows an approximate 1/f
fluctuation with the power index of -0.81 for the frequency range
extended down to $3\times10^{-7}Hz$. This range partially overlaps
the observed range below $2\times10^{-6}$ Hz, although a full description
of the 1/f feature in earthquakes is premature. We want to elaborate
our study in order to further extend the PSD power toward lower-frequencies
as observed, including the finer structure of eigenfrequencies, decay
times, and the deviations of the Earth from spherical symmetry or
the elastic body.

To conclude this section, we highlight the critical role of thresholding
in revealing 1/f fluctuations. As demonstrated, the mere superposition
of the eigenmodes of EFO, denoted as $\Phi(t)$, yields no signal
of 1/f fluctuations in PSD. However, when a threshold is applied to
the continuous data $\Phi(t)$, the resulting thresholded discrete
data reveals prominent 1/f fluctuations, as depicted in Fig. 5. This
thresholded EFO amplitude corresponds to the critical push required
to initiate fault rupture---the primary phase of the earthquake.
Consequently, the occurrence timing of earthquakes displays 1/f fluctuations,
irrespective of the earthquake's energy itself.

In practice, it is acknowledged that not all instances of thresholded
signals successfully lead to earthquake initiation. Nevertheless,
through comprehensive verification, we have generally observed that
any randomly selected subset of data displaying 1/f fluctuations also
exhibits this characteristic in earthquake occurrence timing. This
aligns more closely with realistic earthquake timing occurrences.

Hence, the minimal energy requisite for the critical push to trigger
fault rupture serves as a natural threshold on the raw EFO waves.
This inherent thresholding process unveils the 1/f fluctuation property
in the sequence of earthquake occurrence times.

\section{Discussions \label{sec:Discussions}}

\begin{figure}
\includegraphics[width=12cm]{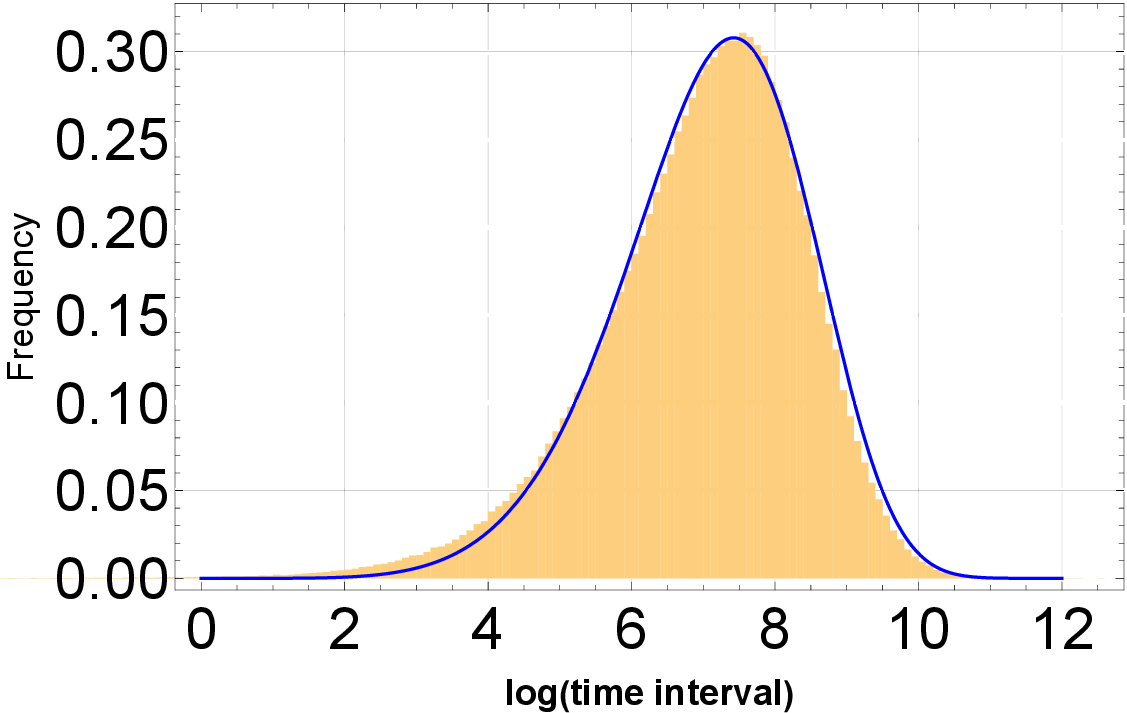}\caption{(Color online) In the graphical representation, the orange bars depict
the frequency distribution of the logarithmic intervals between seismic
activities. The blue line is the Weibull distribution with the parameter
$\alpha=6.3,\beta=7.63$: best fitting the seismic data. }
\label{fig6} 
\end{figure}

In this section, we contrast our 1/f characterization of seismic activity
occurrence sequences with the Weibull distribution \citep{Tanaka2017,Hatano2017}
\begin{equation}
f(x)=\frac{\alpha}{\beta}\left(\frac{x}{\beta}\right)^{\alpha-1}\exp\left(-\left(\frac{x}{\beta}\right)^{\alpha}\right).
\end{equation}
It turns out that the 50-year global seismic activity occurrence time
interval, in logarithm, follows the Weibull distribution, as shown
in Fig.\ref{fig6}.

However, our analysis indicates that time sequences conforming to
the Weibull distribution do not exhibit 1/f fluctuations, as evidenced
by a flat PSD in the low-frequency range. Thus the seismic 1/f fluctuation
is independent of the Weibull distribution. 

This observation can be intuitively grasped. The Weibull distribution
delineates the adjacent occurrence of earthquakes, providing insights
into local properties. Conversely, the 1/f fluctuation characterizes
the low-frequency and long-time correlation among multiple earthquakes,
offering an understanding of global properties. Therefore, it becomes
apparent that the short-term patterns captured by the Weibull distribution
and the long-term correlations described by 1/f fluctuations represent
distinct, yet complementary, aspects of seismic phenomena.

\section{Conclusions and prospects }

We first demonstrated that the time sequence of the magnitude-limited
earthquake energy exhibits 1/f fluctuations. We observed that these
special fluctuations are not apparent when including giant earthquakes
but become more pronounced when energy information is completely excluded.
Therefore, we speculated that the 1/f fluctuation in the earthquakes
is related to the low-energy trigger.

We applied our previous hypothesis suggesting that amplitude modulation,
or the beat of waves with accumulating frequencies, is a common cause
of 1/f fluctuations. A typical mechanism is a resonance. We theorized
that the Earth's lithosphere acts as a resonator, with the Earth Free
Oscillation (EFO) potentially encoding this amplitude modulation.
Then this EFO may trigger the 1/f fluctuation in the earthquake timing
through a tiny final one push toward the fault rupture. This process
corresponds to the demodulation of the encoded 1/f flcutuations.

To test this theory, we constructed data by superposing sinusoidal
waves of the lowest 1158 EFO eigenfrequencies. Then the absolute value
of this time sequence shows 1/f fluctuations with a power index $-1.1$
down to $10^{-4}$Hz.

Additionally, we refined our data to include the resonance effect
and the fine structures denoted by $m$, as induced by Earth's rotation.
We added 1000 extra modes generated by the resonant Lorentzian distributions
for the first 100 eigenfrequencies of EFO after resolving the degeneracy
in m. Then the absolute value of this time sequence shows 1/f fluctuations
with a power index $-0.81$ down to $10^{-7}$Hz. This range partially
overlaps with the observed range of seismic 1/f fluctuations: power
index $-0.8\sim\lyxmathsym{\textendash}1.0$ from about $2\times10^{-6}$
Hz down to $10^{-9}$ Hz. Thus we partially verified that the EFO
triggered seismic 1/f fluctuations.

We aim to expand our research by focusing on more accurately determining
the resonant eigenfrequencies of EFO, incorporating damping effects,
and precisely evaluating excited modes.
\begin{acknowledgments}
We want to acknowledge many valuable discussions with the members
of the Lunch-Time Remote Meeting, the Department of Physics at Ochanomizu
University, and Izumi Uesaka at Kyoto Sangyo\textcolor{red}{{} }University.
This work was supported by JSPS KAKENHI Grant Number 18K18765. 
\end{acknowledgments}


\end{document}